\documentstyle[12pt]{article}

\newcommand{\eq}{\begin{equation}}
\newcommand{\eqn}{\begin{displaymath}}
\newcommand{\en}{\end{equation}}
\newcommand{\enn}{\end{displaymath}}

\def\pnot{\mbox{${\not{\hbox{\kern-3.0pt$p$}}}$}}
\def\qnot{\mbox{${\not{\hbox{\kern-2.0pt$q$}}}$}}

\begin{document}
\begin{titlepage}


\hskip 10cm \vbox{\hbox{CNEA-CAB IT3079/DFUC}\hbox{July 1993}}
\vskip 0.6cm
\centerline{\bf STRINGS AND NON-TOPOLOGICAL SOLITONS$^{\ast}$}
\vskip 1.4cm
\centerline{  R. Fiore$^{\dagger}$}
\vskip .5cm
\centerline{\sl  Dipartimento di Fisica, Universit\`a della Calabria}
\centerline{\sl Istituto Nazionale di Fisica Nucleare, Gruppo collegato di
Cosenza}
\centerline{\sl Arcavacata di Rende, I-87036 Cosenza, Italy}
\vskip 1.2cm
\centerline{  D. Galeazzi, L. Masperi$^{\ddagger}$, A. Megevand}
\vskip .5cm
\centerline{\sl  Centro At\'omico Bariloche and Instituto Balseiro}
\centerline{\sl  (Comisi\'on Nacional de Energ\'{\i}a At\'omica and
Universidad de Cuyo)} \centerline{\sl 8400  S. C. de Bariloche, Argentina}
\vskip 2cm \begin{abstract} We have numerically calculated topological and
non-topological solitons in two spatial dimensions with Chern-Simons term.
Their quantum stability, as well as that of the Maxwell vortex, is analyzed
by means of bounce instantons which involve three-dimensional strings and
non-topological solitons.
\end{abstract} \vskip .5cm
\hrule \vskip.3cm
\noindent

\noindent
$^{\ast}${\it Work supported in part by the Ministero dell'Universit\`a e della
Ricerca Scientifica e Tecnologica of Italy and in part by the Consejo Nacional
de Investigaciones Cientificas y T\'ecnicas of Argentina}
\vfill

$ \begin{array}{ll}
^{\dagger}\mbox{{\it email address:}} &
 \mbox{40330::FIORE}\\
 & \mbox{FIORE@COSENZA.INFN.IT}
\end{array}
$

$ \begin{array}{ll}
^{\ddagger}\mbox{{\it email address:}} &
 \mbox{MASPERI@ARIB51.BITNET}
\end{array}
$
\vfill
\end{titlepage}
\eject
{\bf 1. Introduction}

Gauge models of a scalar field with Chern-Simons term in two spatial
dimensions are an interesting theoretical laboratory for the study of both
topological and non-topological solitons~\cite{J1}. Their existence has
been suggested in the thin-wall approximation beyond the favourable region
of parameters for the potential~\cite{Ma} giving metastable configurations
which provide an inhomogeneous nucleation mechanism for first-order phase
transition. The importance of these solitons depends on their quantum
stability that is affected by bounce instantons which involve the
existence of three-dimensional defects~\cite{PV}.
\vskip.3cm
We have proved by numerical computation the existence of the
two-dimensional solitons with the above anticipated features. At the same
time we have examined the possibility of their quantum decay as well as
that of the Maxwell vortex. This analysis in thin-wall approximation requires
the existence of strings which correspond to a model enlarged with a second
scalar field for which well-known three-dimensional $Q$-balls~\cite{CF} are
possible, producing quantum decays of the two-dimensional solitons.
\vskip.3cm
The $3+1$ generalization of the Chern-Simons theory may be interpreted as
an axion toy model whose strings might give an inhomogeneous nucleation
mechanism for the phase transition of a simplified electroweak symmetry
 breaking.
\vskip.3cm
In section 2 we give a description using thin-wall approximation of the two
and three dimensional defects which will be relevant for our discussion.
Section 3 contains the numerical solutions of two dimensional Chern-Simons
topological and non-topological solitons. Section 4 treats the quantum
decays of Maxwell and Chern-Simons two-dimensional defects using instantons
in thin-wall approximation. Some outlook is contained in section 5.

\vskip 0.5cm

{\bf 2. Defectes of abelian gauge model for scalar fields}

We begin describing in the thin-wall approximation the solitons which
appear for a theory with two scalar fields coupled to independent $U(1)$
gauge fields, which will be relevant for the subsequent discussion. Let us
consider the Lagrangian
\eq
{\cal L} = \left(D_{\mu}\phi\right)^*D^{\mu}\phi+\left(D_{\mu}\sigma\right)^
*D^{\mu}\sigma-V(|\phi|,|\sigma|)+T_{kin}~,
\label{z1}\en
where
\eqn
D_{\mu}\phi = \left(\partial_{\mu}+ig'Z_{\mu}\right)\phi~,~~~~~~~
D_{\mu}\sigma = \left(\partial_{\mu}+igA_{\mu}\right)\sigma
\enn
and it is necessary that the potential has two minima for different values
of $|\phi|$ and $|\sigma|$, being e.g. of the form
\eq
V(|\phi|,|\sigma|) = h\left(|\phi|^2-v^2\right)^2+g|\sigma|^4+f|\phi|^2
|\sigma|^2-M^2|\sigma|^2~.
\label{z2}\en
The kinetic term $T_{kin}$ for gauge fields will be specified below.
\vskip.3cm
If the two minima correspond to
\eq
|\phi| = v~,~~~~~~~~\sigma = 0~
\label{z3}\en
and to
\eq
\phi = 0~,~~~~~~~~|\sigma| = \sigma_0~
\label{z4}\en
for the case that the stable vacuum is represented by eq.(\ref{z3}), in two
spatial dimensions one may build a vortex with $\phi$ such that
\eq
\phi{\rightarrow}ve^{i\theta}~,~~~~~~~~\sigma{\rightarrow}0~~~~~~{\rm
for}~~~r{\rightarrow\infty}
\label{z5}\en
together with an angular component of gauge field $Z_{\theta}$ to make the
energy finite. To minimize the energy it is convenient that in the vortex
core $\phi\sim 0$,~$\sigma\sim \sigma_0$.
\vskip.3cm
Choosing the Maxwell kinetic term for the relevant gauge field:
\eq
T^M_{kin} = -{1\over 4}\left(\partial_{\mu}Z_{\nu}-\partial_{\nu}Z_{\mu}
\right)^2~,
\label{z6}\en
the vortex energy, or energy per unit string length if the third spatial
dimension $z$ is added, for static configurations will be
\eq
E_v = aR^2+bR+{c\over{R^2}}
\label{z7}\en
in terms of core radius R. Here the coefficients $a$, $b$ and $c$ are
positive, the first term comes from the false vacuum region, the second one
from the interface contribution whereas the third one arises from the
magnetic energy with the constraint of constant flux ${\pi}R^2B$.
\vskip.3cm
The minimization of eq.(\ref{z7}) with respect to $R$ gives a classically
stable configuration. If we additionally consider excitations of the core,
\eqn
\sigma = exp[i\nu(z,t)]\sigma_0~,
\enn
where also the second gauge field $A$ will be present, which we may take as
the electromagnetic one, the superconducting cosmic strings~\cite{Wi}
emerge stabilized by constancy of the phase change along a closed string:
\eq
\oint dl{d\nu\over{dl}} = 2{\pi}N~.
\label{z8}\en
\vskip.3cm
For the Lagrangian (\ref{z1}) without gauge fields and with stable vacuum
(\ref{z3}) it is also possible to have classically stable configurations of
real $\phi$ in three spatial dimensions where, inside a localized region
 $\phi\sim 0$, $\sigma\sim \sigma_0e^{-i{\omega}t}$. Now, again in the
thin-wall approximation, the energy will have the radius dependence
\eq
E_{\sigma} = aR^3+bR^2+{c\over{R^3}}~,
\label{z9}\en
where the last term comes from the kinetic term of $\sigma$ with the
condition of fixed charge. The minimization of the energy with respect to
$R$ produces a non topological soliton called $Q$-ball which is proved to be
stable for a charge larger than a threshold~\cite{CF}.  \vskip.3cm If the
parameters of the potential (\ref{z2}) are changed so that the minimum
(\ref{z4}) is the absolute one, one has $a<0$ and the string and the
$Q$-ball, in case they survive, become bubbles of the corresponding stable
phase in the metastable sea characterized by eq.(\ref{z3}).  Therefore, they
may give way to an inhomogeneous nucleation mechanism for the transition
from the $|\phi|=v$ phase to the $\phi=0$ one as noticed for Maxwell
vortices~\cite{StJ}.  \vskip.3cm
Taking the Lagrangian (\ref{z1}) with a complex $\phi$ coupled to $Z_{\mu}$
and a real $\sigma$ new types of defects may appear if we replace the
Maxwell kinetic term in $3+1$ dimensions by \eq
T^{C-S}_{kin} = {1\over 2}\gamma\varepsilon^{\mu\nu\rho\tau}\partial_{\mu}
{\sigma}Z_{\nu}\partial_{\rho}Z_{\tau}~,
\label{z10}\en
where notations will be justified below.
The energy for static configurations will be
\eq
E= \int d^3x\left[|\vec{D}\phi|^2+(\vec{\nabla}\sigma)^2+V+{\gamma^2\over
4}{(\vec{\nabla}\sigma\cdot\vec{B})^2\over{|\phi|^2}}\right]~,
\label{z11}\en
where minimization on $Z_0$ has been performed.
\vskip.3cm
We must note that in $2+1$ dimensions $\vec{B}=B_z$ and $\partial_z\sigma$
may be replaced by a constant. In this way the problem described by eq.s
(\ref{z10}) and (\ref{z11}) reduces to that with Chern-Simons term
which has been studied~\cite{J1} for a sextic potential in $\phi$ without
$\sigma$ and with two degenerate minima showing
the existence of topological vortices for $|\phi|\rightarrow v$
and non topological solitons for $\phi\rightarrow 0$ when $r\rightarrow
\infty$. The latter are possible, at variance from the Maxwell case, because
the last term of eq.(\ref{z11}) requires that $B$ is confined in the region
where $\phi$ is nonvanishing. To pass from the potential (\ref{z2}) to an
effective one $V_{eff}(\phi)$ depending only on $\phi$ formally requires a
minimization with respect to $\sigma$. This shows that for $|\phi|\sim v$,
if the reasonable condition $fv^2>M^2$ is satisfied, $v$ is a minimum of
$V_{eff}=0$, whereas for $\phi\sim 0$ $V_{eff}(0)<V_{eff}(v)$ if
$hv^4<{M^4\over{4g}}$.
\vskip.3cm
The analysis of these two-dimensional solitons in the thin-wall
approximation shows~\cite{Ma} that topological vortices are favoured when
$|\phi|=v$ corresponds to the absolute minimum but that they survive when
$\phi=0$ becomes the true vacuum up to a critical difference between both
minima. Conversely the non-topological solitons prefer the situation where
$\phi=0$ is the absolute minimum but survive as metastable states when the
broken symmetry phase becomes stable up to a critical difference with the
symmetric false vacuum. Therefore, these non-topological solitons provide
an inhomogeneous nucleation mechanism for the first-order transition from
the "over-frozen" symmetric phase to the stable broken-symmetry one when
the potential energy difference is larger than the critical value and the
solutions become classically unstable increasing their radius indefinitely.
The existence of these solutions has been confirmed numerically as will be
discussed in the next section.
\vskip.3cm
Returning to the 3+1 dimensional case, the generalization of two
dimensional solitons is not immediate as in the case of the Maxwell string,
because $\partial_z\sigma$ must be nonvanishing but cannot be constant for
all values of $r$. Therefore interesting solutions are $z$-dependent
configurations which, due to the equations for gauge field $Z_{\mu}$
coupled to the current $J_{\mu}$:
\eq
{\gamma\over 2}\vec{\nabla}\sigma\cdot\vec{B} = J_0~,~~~~~~{\gamma\over
2}\vec{\nabla}\sigma\times\vec{E} = \vec{J}~,
\label{z12}\en
together with the additional radial dependence of $\sigma$, $Z_\theta$ and
$Z_0$, give charge and angular $J_{\theta}$ current. If we wish $\sigma\sim
0$ in the core and $\sigma\sim\sigma_0$ outside, charge and current will be
concentrated on the string surface.
\vskip.3cm
To avoid values of $z$ for which the string size vanishes it is necessary
that $\sigma $ is an increasing function of $z$. This is possible if
$\sigma$ is originally a phase as occurs in the Peccei-Quinn axion model
{}~\cite{PQ}. In this case the renormalizable potential (\ref{z2}) must be
replaced by e.g.
\eq
V(|\phi|,|\sigma|) = h\left(|\phi|^2-v^2\right)^2+\left[m^2_af^2_{PQ}+
k\left(|\phi|^2-v^2\right)\right]\left(1-cos{\sigma\over f_{PQ}}\right)
\label{z13}\en
which has a similar form for small $\sigma$ if $kv^2>m^2_af^2_{PQ}$.
Eq.(\ref{z13}) corresponds to the axion effective potential~\cite{VW} when
$|\phi|=v$ whereas for $\phi=0$ has its minimum for $\sigma/f_{PQ}=\pi$.
This may be qualitatively interpreted by the fact that before the
electroweak symmetry breaking transition there is no contribution of
$arg$ $det\cal M$, being $\cal M$ the fermion mass-matrix, to the strong $CP$
violating parameter.
\vskip.3cm
To understand the non-topological string in the thin-wall approximation one
must consider the width $\Delta$ around an average radius $R$ in which the
transition between the core at $\sigma=0$ and the outside region at $\sigma
/f_{PQ}=\pi$ may occur. Moreover one has to keep in mind that the true
variable is $U=\exp(i{\sigma\over{f_{PQ}}})$
which passes from $1$ to $-1$. As indicated in Fig.1,
starting from a value of $z$ for which the transition occurs just below
$R+{\Delta\over 2}$, $\left(a\right)$, increasing $z$ we lower the transition
point almost down to $R-{\Delta\over 2}$, $\left(b\right)$. At this moment
for the remaining very small region still at $\sigma/f_{PQ}=0$ there is a
jump to $2\pi$ $\left(c\right)$. Then the transition point moves to the
right and when it almost reaches $R+{\Delta\over 2}$, $\left(d\right)$, the
remaining small region still at $\pi$ jumps to $3\pi$ completing the cycle
and returning to $\left(a\right)$.  \vskip.3cm For a closed string, along
its length $L$ an integer number of cycles must be performed which gives a
sort of topological stability to the configuration.  In first approximation
the average radius of the core can be evaluated as in the two-dimensional
case~\cite{Ma} replacing the parameter $\mu$ of the Chern-Simons term by
$\gamma\partial_z\sigma\approx\gamma f_{PQ}\cdot2\pi{N\over L}$. It is clear
that the two discontinuities per cycle described above produce divergences
of $\partial_z\sigma$ which affect the second and last terms of the
integrand of eq.(\ref{z11}). This, however, gives only finite contributions to
the energy due to the vanishingly small regions of $r$ and $z$ in which
they appear. The jumps in $\sigma$ are only possible as quantum effects
which must be penalized by probability factors as occurs in cosmic strings
for abrupt change of phase~\cite{Wi}. These jumps are the most economic ones
for what concerns the energy of the configuration and allow the building of
the non-topological string with only a gentle oscillation of transverse size
around $R$ going along its length.

\vskip 0.5cm

{\bf 3. Numerical solutions of Chern-Simons solitons}

As indicated in the previous section, the two-dimensional Chern-Simons
topological and non-topological solitons may provide an inhomogeneous
nucleation mechanism for the transition to the symmetric or to the
symmetry-broken phase respectively.
\vskip.3cm
For that it is important to prove numerically the existence of the solution
outside the favourable region of the potential parameters. Taking the
sextic potential
\eq
V_{eff} = {1\over{\mu^2}}\left(|\phi|^2-v^2\right)^2\left(|\phi|^2+\lambda
v^2\right)
\label{z14}\en
which has two minima for $-1<\lambda<{1\over 2}$, and representing the
relevant fields as
\eq
\phi(r,\theta) = {v\rho(r)}e^{in\theta}~,~~~~~~Z_{\theta}(r) = {1\over r}
[n-\zeta(r)]~,
\label{z15}\en
the static classical solutions must satisfy~\cite{Ba}
\eqn
{{d^2\zeta}\over{dx^2}}-{1\over x}{d\zeta\over{dx}}-{2\over\rho}{d\rho
\over{dx}}{d\zeta\over{dx}}-{{4{\rho}^2\zeta}\over{{\mu}^2}} = 0~,
\enn
\eq
{{d^2\rho}\over{dx^2}}+{1\over x}{d\rho\over{dx}}+{\mu^2\over{4x^2\rho^3}}
\left({d\zeta\over{dx}}\right)^2-{1\over
x^2}{\zeta}^2\rho-\rho\left(\rho^2-1\right) \left(3\rho^2+2\lambda-1\right)
= 0~, \label{z16}\en where $x=v^2r$ and we have included the coupling
constant $g'$ into $Z_ {\theta}$.
\vskip.3cm This highly non-linear system of equations has been numerically
solved for the topological
\eqn \rho = 1~,~~~~~\zeta = 0~~~~~{\rm for}~~r\rightarrow\infty~,
\enn \eq
\rho = 0~,~~~~~\zeta = n~~~~~{\rm for}~~r\rightarrow 0~ \label{z17}\en
asymptotic behaviours taking $n=1$, and for the non-topological ones \eqn
\rho = 0~,~~~~~\zeta = -\alpha~~~~~{\rm for}~~r\rightarrow\infty~, \enn
\eq \rho = 1~,~~~~~\zeta = 0~~~~~{\rm for}~~r\rightarrow 0
\label{z18}\en taking $n=0$ and $\alpha=-0.8$, choosing in all cases
$\mu^2=1$.  \vskip.3cm
The natural region for the topological solutions corresponds to $\lambda>0$
where the absolute minimum of $V_{eff}$ is at $|\phi|=v$. But solutions
appear also for $\lambda<0$, the criterium being that $B$ decreases for
large values of $x$ where $\rho$ has already attained its asymptotic limit
in order to ensure finiteness of energy. Acceptable and non acceptable
solutions for $\lambda<0$ are shown in Fig.2 $\left(a\right)$ and
$\left(b\right)$ respectively. From the numerical results we may establish
that $-0.1>\lambda^{T}_{crit}>-0.2$.
\vskip.3cm
On the other hand the natural region for non-topological solitons is
$\lambda<0$ where the absolute minimum of $V_{eff}$ is at $\phi=0$.
Acceptable solutions appear also for $\lambda>0$ (Fig.3$\left(a\right)$)
up to a critical value beyond which incorrect behaviours for $\zeta$ and $B$
at small $x$ force us to reject them (Fig.3$\left(b\right)$) which allows us
to estimate that $\lambda^{NT}_{crit}\simeq 0.14$.

\vskip 0.5cm

{\bf 4. Quantum stability of Maxwell vortices and Chern-Simons non-
topological solitons}

The quantum stability of two-dimensional solitons is related to the
three-dimensional ones which appear as bounce type instantons in
Euclidean metric
{}~\cite{PV} giving a decay probability $e^{-S}$ with $S$ bounce action taken
as difference between instantons and unperturbed string actions.
\vskip.3cm
In order to compare with our Chern-Simons solitons we start from what
occurs to Maxwell vortices. Apart from the possibility of a
monopole-antimonopole pair produced at higher scale if a hierarchy
symmetry exists
which would allow an intermediate state without flux, another channel
corresponds to a bounce which includes $Q$-balls existing in the same
generalization to 3+1 dimensions of the model. In fact for Lagrangian
(\ref{z1}) with potential (\ref{z2}), if the true vacuum corresponds to
eq.(\ref{z3}) for which the vortex is favoured, the three-dimensional string
may end on a $Q_{\sigma}$-ball of mass $E_{\sigma}$ made by the charge of
field $\sigma$ which is stable in this case. Since beyond it the
intermediate configuration has $\phi=v$, the magnetic flux is dispersed in
an unconfined region of vanishingly small field (Fig.4$\left(a\right)$) that
corresponds to the broken-symmetry vacuum to which the vortex may decay.
The bounce action will be estimated as the mass of the $Q_{\sigma}$-ball
pair and its Coulomb interaction
\eq
S = 2E_{\sigma}- {Q_{\sigma}^2\over l}-E_vl~,
\label{z19}\en
where the last term is the missing string energy with $E_v$ energy per unit
length, and the unstable maximum with respect to $l$ must be taken.
\vskip.3cm
If the potential is such that eq.(\ref{z4}) corresponds to stable vacuum,
the possible $Q$-ball is that inside which there is the charge
corresponding to field $\phi$. This defect, where fields $\phi$ and
$\sigma$ are exchanged respect to the discussion of section 2, will be
denoted as $Q_{\phi}$-ball. Now the vortex is clearly metastable so that the
needed instanton requires a small $Q_{\phi}$-ball acting as an obstacle that
deviates slightly the magnetic flux (Fig.4$\left(b\right)$) and giving an
intermediate configuration of larger size which corresponds to the state
reached by tunnel effect through the wall of eq.(\ref{z7}). The bounce
action is similar to that of eq.(\ref{z19}) but the mass $E_{\phi}$ of the
$Q_{\phi}$-ball is smaller and smaller for deeper minimum of eq.(\ref{z2})
at $\phi=0$, making the decay more probable.  \vskip.3cm

For the Chern-Simons non-topological soliton in two dimensions the
potential (\ref{z13}) must be taken to allow the extension to three
dimensions. Now the situation is somehow reversed compared to the Maxwell
vortex. When the stable vacuum corresponds to eq.(\ref{z4}) the stability
of non-topological soliton is favoured, the only instanton which allows its
decay being that where the string ends in a $Q_{\phi}$-ball which disperses
the flux since magnetic field must vanish for $\phi=0$
(Fig.5$\left(a\right)$). On the contrary, when the stable vacuum becomes that
of eq.(\ref{z3}), the obstacle of a small $Q_{\sigma}$-ball (\ref{z2})
inside which $\phi=0$ enlarges the size of the defect(Fig.5$\left(b\right)$)
in correspondence to tunnel effect from the metastable initial state.

\vskip 0.5cm

{\bf 5. Conclusions}

We have numerically shown the existence of two-dimensional Chern-Simons
solitons also outside their favoured range of potential parameters,
confirming what anticipated by the thin-wall approximation. The faster decay
of the metastable case has been understood by the extension of the model to
3+1 dimensions. The three-dimensional non-topological string appears in a
model which can be understood as that of a scalar Higgs field in
interaction with an electrically neutral gauge field additionally coupled
to an axion field. Compared to the standard model, the $SU(2)$ gauge
coupling constant has to be put equal to zero, apart from taking no care
about the phenomenological values of the axion sector.
\vskip.3cm
Making the model more realistic and including temperature effects, studying
numerically the three-dimensional non-topological strings and analyzing
their quantum stability it would be possible to compute their relevance as
an inhomogeneous nucleation mechanism for the phase transition of the
electroweak symmetry breaking.

\vskip 1.5cm
{\bf Acknowledgement}:We are indebted to A.Quartarolo for his
collaboration in technical aspects of the manuscript. L.M. wishes to thank
the hospitality at the Dipartimento di Fisica dell'Universit\`a della
Calabria where part of this work has been performed.

\newpage

\centerline{\bf Figure Captions}
\vskip .3 cm
\begin{description}

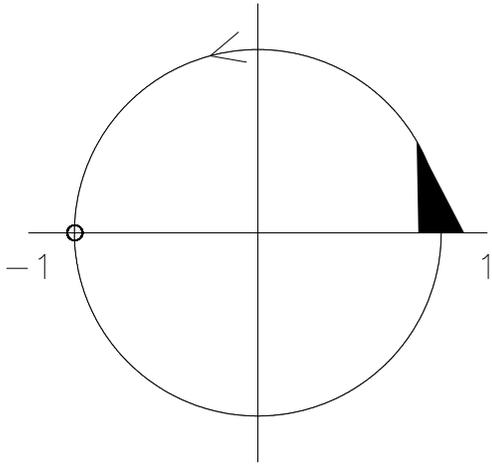
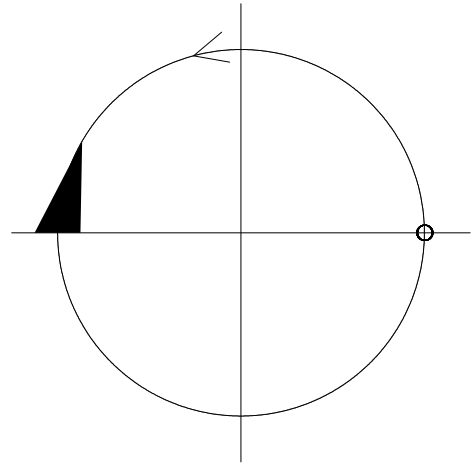
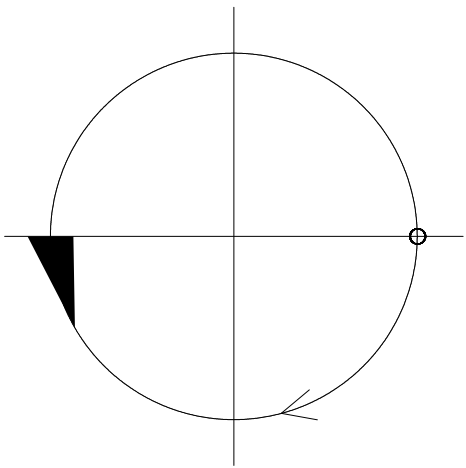
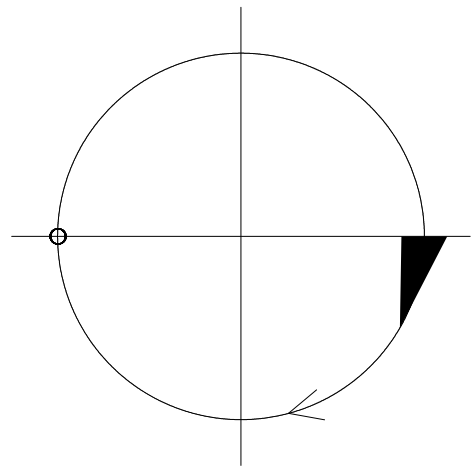
\item{Fig.1:}
Evolution of U along the non-topological string. Arrows indicate
mapping of $R-{\Delta\over 2}<r<R+{\Delta\over 2}$. Triangles show values
of U taken by most of the radial interval.
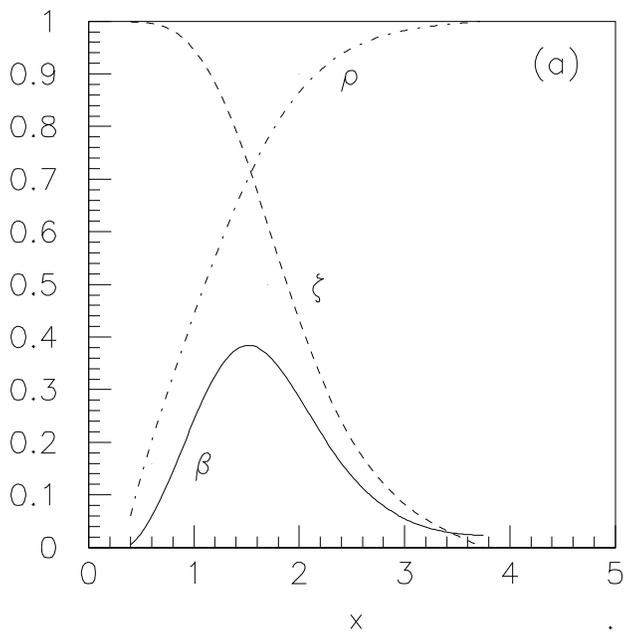
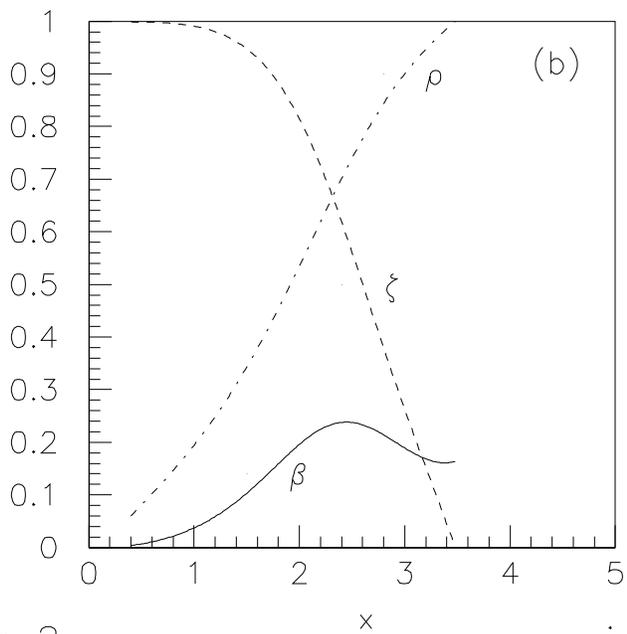
\item{Fig.2:}
Numerical solution, (a) acceptable ($\lambda=-0.05$) and (b) non-
acceptable ($\lambda=-0.25$), for the two-dimensional Chern-Simons topological
vortex.
We define $\beta={B\over {v^4}}$ and $x=rv^2$.
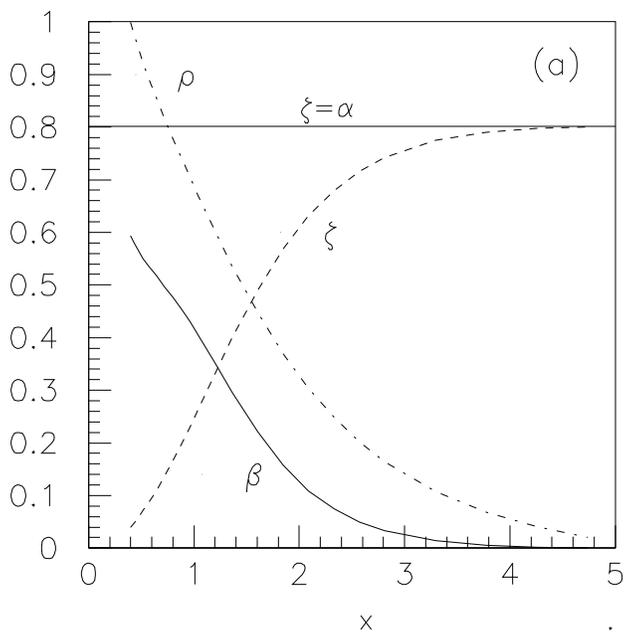
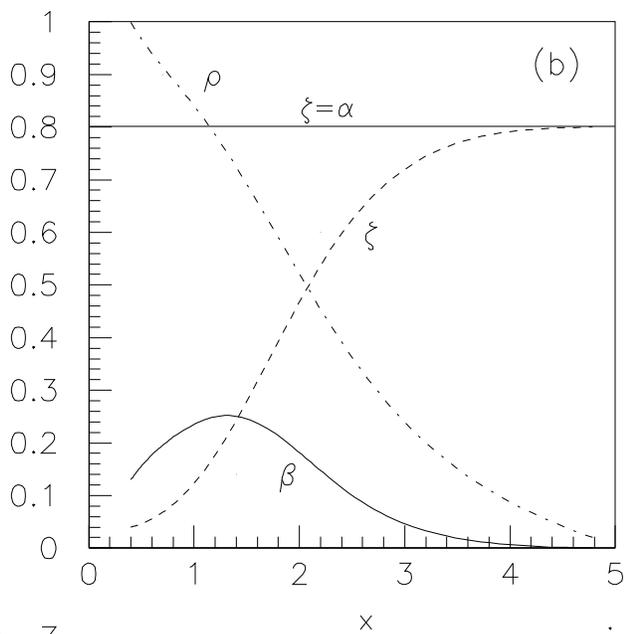
\item{Fig.3:}
Numerical solution, (a) acceptable ($\lambda=0.1$) and (b) non-
acceptable ($\lambda=0.2$), for the two-dimensional Chern-Simons
non-topological soliton.
We define $\beta={B\over {v^4}}$ and $x=rv^2$.
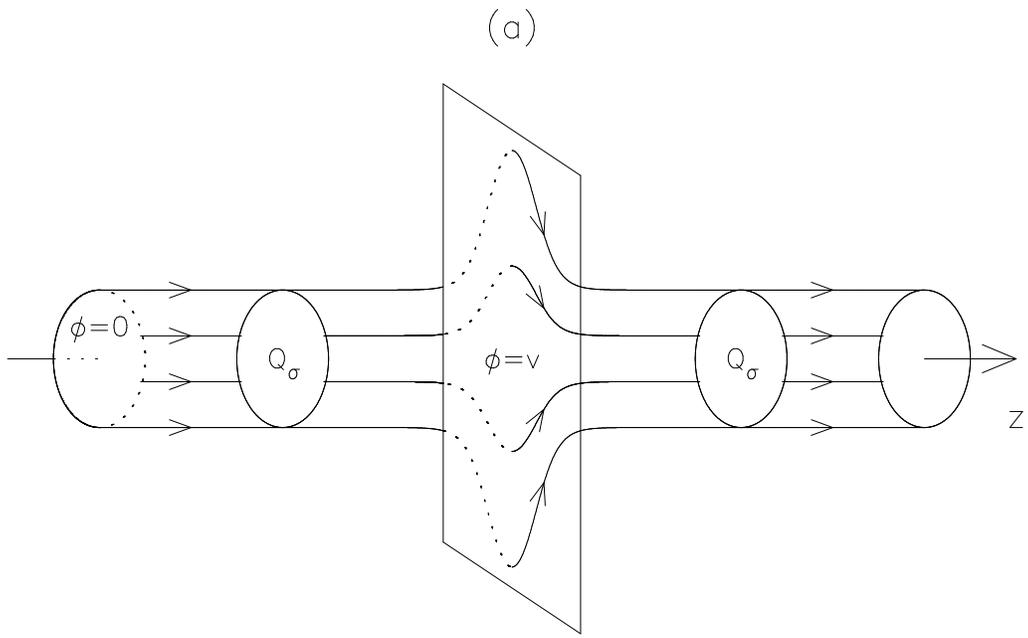
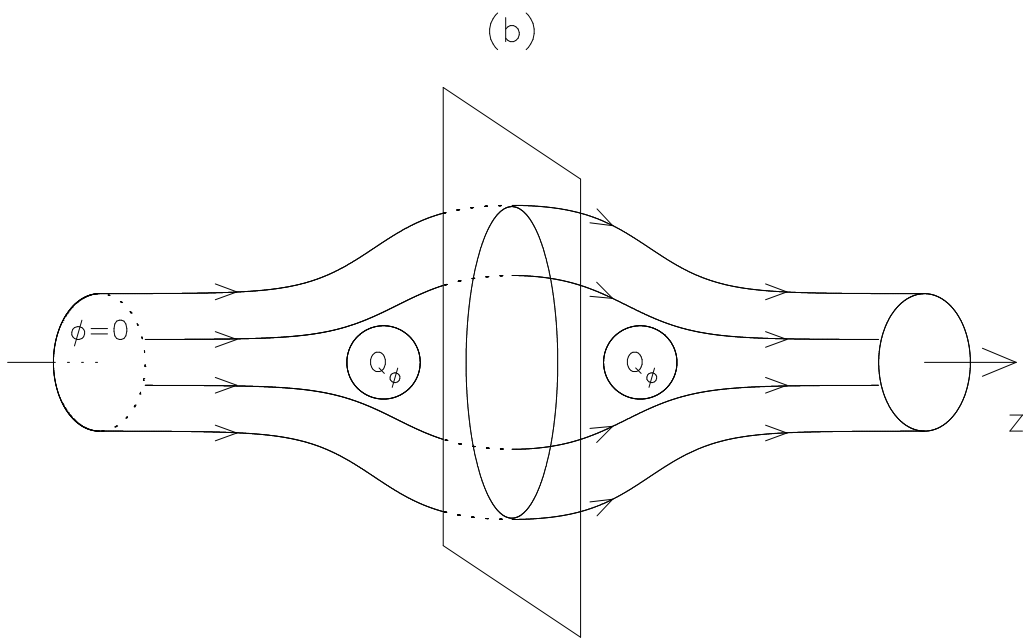
\item{Fig.4:}
Instanton describing the decay of the two-dimensional Maxwell vortex.
In case (a) the stable vacuum corresponds to $|\phi|=v$, $\sigma=0$.
The equivalent string ends on $Q_{\sigma}$-ball. The intermediate
configuration indicates the decay to the vacuum.
In case (b) the stable vacuum corresponds to $\phi=0$, $|\sigma|=\sigma_0$.
The equivalent string finds obstacle of $Q_{\phi}$. The intermediate
configuration indicates the decay to a broader vortex.
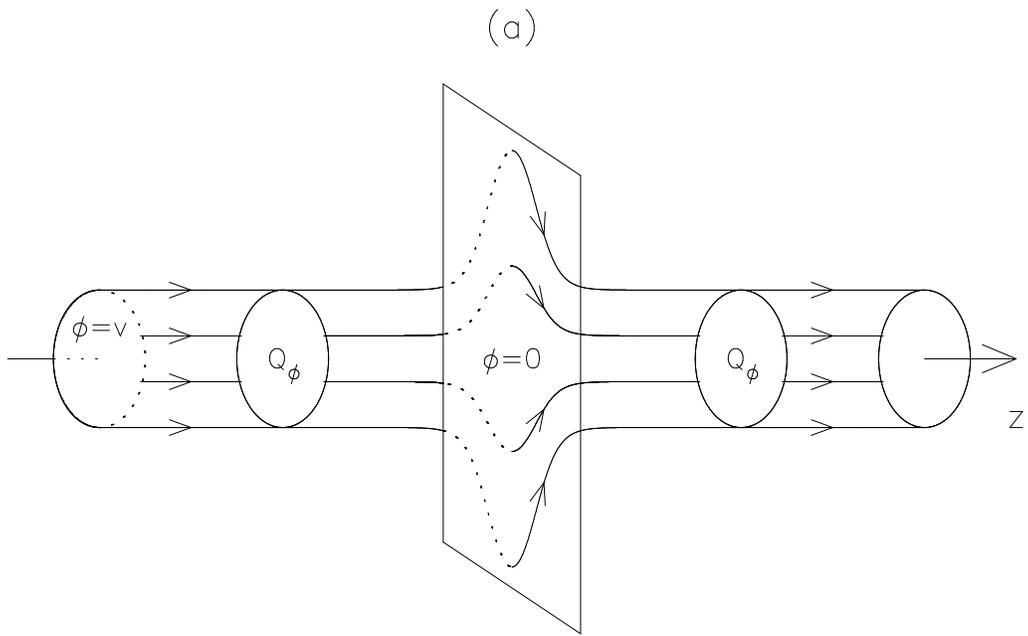
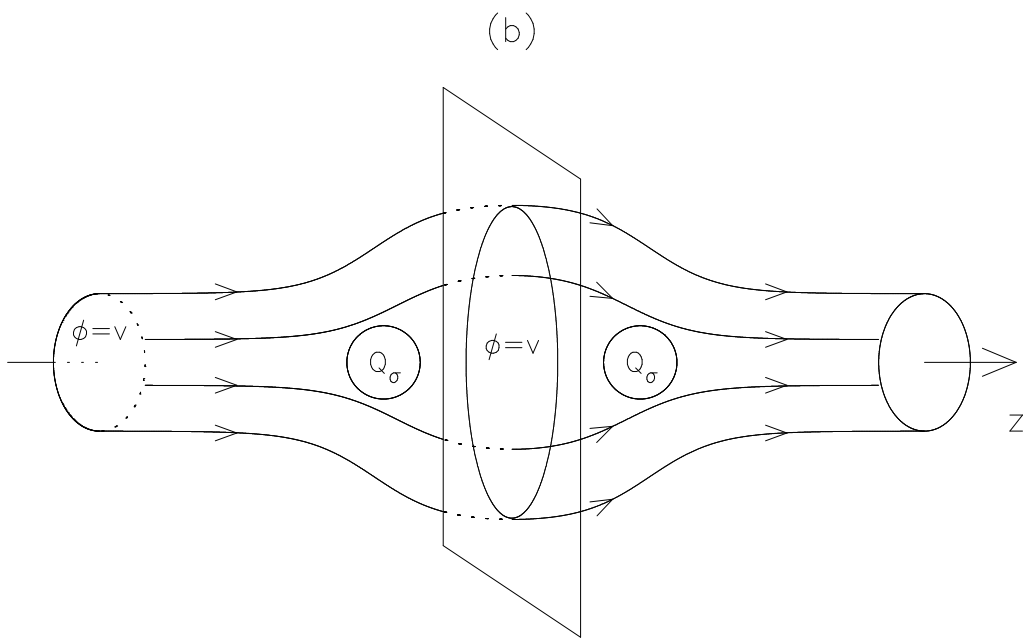
\item{Fig.5:}
Instanton describing the decay of the two-dimensional Chern-Simons
non-topological soliton.
The stable vacuum corresponds (a) to $\phi=0$, $|\sigma|=\sigma_0$,
(b) to $|\phi|=v$, $\sigma=0$.

\end{description}

\end{document}